\documentclass[11pt,twoside]{article}  
\usepackage{apn3conf}
\begin{document}   

%
%
%
%

\title{Chemical Abundances and Their Relation with PN Morphological Features}

%
%
%

\author{Antonio Mampaso}
\affil{ Instituto de Astrof{\'{\i}}sica de Canarias,
C/V{\'{\i}}a L\'actea s/n, 38205 La Laguna, Tenerife, Spain}

%
%

\contact{A Mampaso}
\email{amr@iac.es}

%
%
%
%
%

\paindex{Mampaso, A.}

%
%

\authormark{Mampaso}

%
%


\begin{abstract}          
Results appeared since the last APN meeting on the relation between
abundances and PNe morphological features are reviewed here.
Statistical analyses of round, elliptical, and bipolar PNe clearly
separate the latter group as having larger He and N/O abundances.
Studies of individual PNe, including the UV and IR domains, allow a
detailed comparison with evolutionary models for the AGB phase; the
abundances of most important elements can be matched but an
'age-metallicity' problem for the progenitors of He-rich PNe is implied.
Halo abundances do not show strong chemical differences with respect to
the PN rims, but the halos' faintness requires deeper data to confirm it. 
Data on the chemistry of microstructures of
normal (non H-deficient) PNe are scarce and generally indicate similar
abundances than in the rims. The exception is the high nitrogen
overabundance claimed for FLIERS; preliminary modeling on new
observational data of the microstructures of NGC~7009 indicates that a
moderate enrichment of nitrogen can be real.
\end{abstract}

%
%

\section{Introduction }
Planetary nebulae (PNe) arise from the evolution of the most abundant stars (1--8 $M_{\odot}$), and they are a 
key to
understand the chemical evolution of galaxies and the production of crucial elements such as N and C. The gas 
in currently observed PNe 
originated only a few thousand years ago from the atmospheres of AGB stars but the information concerning its 
physical properties 
(density, temperature, kinematics, etc.), which are all very different now from when it belonged to the star, 
has been irretrievably lost. However, the PN
gas does preserve the chemical information: a few elements change during the evolution of low and intermediate 
mass 
stars (notably He, C, N and s-elements), and we should be able to predict how they change, while most elements
do not appreciably change (Ar, S, Ne, O, Cl, etc.), providing us with information on their pristine abundances 
at the
time when the progenitors were formed. 
The chemical history of
the stars is therefore attainable, although it is not at all easy to decipher. One of the great challenges 
in modern
astrophysics is to link the abundances that we observe today in PNe with a stellar evolution theory that 
includes 
the (still poorly known) details of nucleosynthesis, convection plus dredge-up processes, and mass loss, all 
complicated functions of progenitor mass, metallicity, and time. 

Chemical abundances are also a tool for investigating how and in what evolutionary phases the
PN morphological features originate (the main shells, the halos, and the so -called 
microstructures), and this information is in turn
a key to understanding how the PNe themselves were formed. PN global abundances, halo abundances, and 
the still poorly known chemical composition of the microstructures will be discussed here in Sections 2, 3, and 
4, 
respectively. Each one sheds light
on the chemical history of the stars and on the origin of the structures in PNe.

\section{Abundances and Main Shell Morphology}
Since the work of Manuel Peimbert and colleagues in the seventies, it has been noticed that Type I PNe 
(defined as having high He and N abundances) often show bipolar morphology, and quests began for correlations 
between chemical abundances and PN morphology in the hope that they would provide i) clues to the  origin 
of the morphology \footnote {Note that we do not have yet a theory that explains if and why a, say, 1 
$M_{\odot}$ star 
will form a bipolar, an elliptical, or a round PN.} and ii) information on the masses and ages of the 
progenitor stars. 
Pottasch (2000) 
in his review at the last APN 
conference, presented accurate abundances for 16 PNe, concluding that `no simple relationship between 
morphology 
and composition exists.' Two recent papers by Manchado (2003) and Phillips (2003) analyze abundances of some 
75--85 PNe taken from the literature (Phillips 2003) or recalculated from published spectral data using the 
``constant $T_{\rm e}$, $n_{\rm e}$" method and 
IRAF routines (Manchado 2003). Table ~\ref{T1.10-tbl-1} shows a summary of their results for He and the N/O 
ratio; 
data from Perinotto \& Corradi (1998) for a sample of 15 bipolar PNe have also been included for comparison. 
Note 
that values after the $\pm$ sign are 
not errors in quoted abundances but are instead 1-$\sigma$ dispersions of the sample (Phillips 2003 and 
Perinotto 
\& Corradi 1998) and the 50\% percentile (Manchado 2003). Typical uncertainties in individual measures are 
difficult to 
estimate but probably range from 5 to 15\% for He and 30--300\% for metals. The table shows a small marginal 
increase in He 
from round, to elliptical, to bipolar PNe, plus a very high increase in N/O for bipolar PNe.  This reinforces 
the 
current view that considers round PNe as originating from long-lived low-mass stars, whereas bipolar PNe would 
come 
from more massive and younger progenitors. Elliptical PNe would span a wide range of masses and ages (Phillips 
2003). 
Given the uncertainties mentioned above, it is difficult to extract further trustworthy information from such
statistical correlations, and one should move toward a better nebular analysis of individual 
objects. UV, optical, and IR (ISO) data now allow 30\% accuracy in abundances for C, N, O, Ne, S, and Ar, their 
ionization correction factors (ICF) being around unity (Pottasch 2000). Helium abundance accuracy is around 
5\%. Abundances have recently been 
calculated by different authors
in this way for some 20 PNe, although targets are necessarily biased towards 
bright and nearby (0.5 to 2.5 kpc) objects.

\begin{deluxetable}{crrrc}
\scriptsize
\tablecaption{PNe abundances and morphological classes. \label{T1.10-tbl-1}}
\tablehead{\colhead{}& \colhead{Round}& \colhead{Elliptical} & \colhead{Bipolar} & \colhead{{References}}} 
\startdata
He/H &  0.102$\pm$0.010  & 0.121$\pm$0.015  & 0.136$\pm$0.010 & (a) \nl
      & 0.107$\pm$0.010  & 0.115$\pm$0.014  & 0.132$\pm$0.015 & (b) \nl
 & $-$  & $-$  & 0.150$\pm$0.040 & (c) \nl
N/O & 0.27$\pm$0.08  & 0.31$\pm$0.15  & 1.30$\pm$0.50 & (a) \nl
    & 0.22$\pm$0.06  & 0.33$\pm$0.16  & 0.90$\pm$0.21 & (b) \nl
 & $-$  & $-$  & 1.39$\pm$1.02 & (c) \nl
\enddata
\tablenotetext{}{(a) Manchado (2003); (b) Phillips (2003); (c) Perinotto \& Corradi (1998)}
\end{deluxetable}

A brief summary of the best-observed PNe is given in below, together with a comment on the global results for 
each
morphological class. {\it {Round PNe}}: A39 (Jacoby et al.\ 2001), IC~2165 and NGC~5882 (Pottasch et al.\ 
2003a). These PNe
show less processing (He, N, and C) than more structured PNe and would come from low mass 
progenitors having 
metallicities from solar to 2--3 times solar. {\it {Elliptical PNe}}: NGC~7662, NGC~6741 (Pottasch et al.\ 
2001), NGC~40, 
NGC~6153 (Pottasch et al.\ 2003b), and NGC~6543 (Hyung et al.\ 2000; Bernard-Salas  et al.\ 2003). Some of 
these PNe
show very little processing (NGC~7662) and others much more (NGC~6543). Again they should come from low mass 
progenitors with different metallicities, from half 
solar to 2--3 times solar. {\it {Bipolar PNe}}: NGC~6302, NGC~6445, NGC~6537, He2-111 (Pottasch et al.\ 2000), 
NGC~7027 
(Bernard-Salas et al.\ 2001) and Hb-5 (Pottasch et al.\ 2003c). The progenitors are all of intermediate mass, 
3-4 $M_{\odot}$ 
(NGC~7027, 
NGC~6445, and possibly Hb-5) to larger than 4 M$_{\odot}$ (NGC~6302, NGC~6537, and He2-111). These PNe show 
large He 
and N processing.

With accurately determined abundances, a sensible test for models is possible. Marigo et al.\ (2003) compare 
data
for ten PNe with surface stellar abundances from evolutionary models for the thermal pulse--AGB 
phase.
Two classes of PNe are found according to their He abundance; those with He/H $\ge$ 0.15 show much less C and 
O than PNe of lower He/H, and models with different initial conditions are required. The latter group of PNe 
would 
originate from 
AGB stars of low and intermediate mass (0.9--4 $M_{\odot}$) with initial solar metallicity, and their 
abundances of He, N 
and C are well explained with first, possibly second, and third dredge-up processes. The group with He/H $\ge$ 
0.15 would 
instead come from intermediate mass stars (4--5 $M_{\odot}$) of subsolar (LMC-like) metallicity. 
Their abundances (including the low O values) can be explained with both the third dredge-up and the HBB, but 
a strong third dredge-up has to be assumed that very efficiently produces He, but is peculiar in not producing 
much C. 

In summary, Marigo et al.\ (2003) are able to explain the observed PN abundances but find an important 
``age-metallicity'' problem: 
if PNe with He/H $\ge$ 0.15 come from high mass 
progenitors then these should have formed relatively recently, and this seems to be in conflict with their 
having a low (LMC) 
metallicity (the Galaxy is expected to increase secularly in metallicity). Either some physical problem has not 
been 
properly taken into account or there are important chemical peculiarities in the Galaxy. Alternatively, 
abundances in 
high He/H  
PNe (all of them being bipolar) could be affected by binarity.

\section{Halos}
By comparing the abundances in the halos and the main shells of PNe one aims to get information concerning 
when and through what processes the halos
were formed. Guerrero \& Manchado (1999) found similar 
abundances in the halos and the rim of six PNe, whereas in two nebulae, NGC~6720 and NGC~5882, the N$^+$/O$^+$ 
ratio 
in the halo was around half than at the rim. According to those authors, the mild (or null) enrichment 
at the rims is a sign of a small (or null) effect of the third dredge-up in the progenitors. However, the 
$T_{\rm e}$ 
could not be measured for the halos, and a deeper study is desirable.
On the theoretical side, Villaver et al.\ (2002, 2003) find that the ISM can contaminate genuine halos adding 
up 
to 70\%  (for a 1 $M_{\odot}$ progenitor) to the halo mass during the evolution of the nebula. Furthermore, if 
the 
AGB progenitor moves with a velocity of 20 km s$^{-1}$ with respect to the ISM, most (again up to 70\%) of 
the mass ejected could be lost through ram-pressure stripping caused  by the interaction. Both effects ought to 
change the actual halo abundances substantially, thereby complicating interpretations, although no detailed 
predictions 
are available yet.

\section{Microstructures}

\subsection{Cometary Knots}
Their chemistry is not known. O'Dell et al.\ (2000) suggest that [OIII]/H$\alpha$ and 
[NII]/H$\alpha$ are larger in a couple of inner knots than in the outer ones in NGC~7293. This could be caused 
by O and 
N being more 
abundant in the inner regions or (perhaps more plausibly) having a higher $T_{\rm e}$.  More work is needed 
here.

\subsection{Knots in H-Deficient PNe}
A78 and A30 are the best examples of PNe with large abundance variations. Medina \& Pe\~na (2000) confirm 
the extremely high He abundances in the inner knots of A78 but find normal O, N, and Ne, which suggests that 
the gas 
was ejected from zones at the central star after nucleosynthesis took place. They also discovered a 
high velocity outer 
knot, colliding with the shell, where H is underabundant.  This feature deserves a more complete 
chemical 
analysis.

Ercolano et al.\ (2003) modeled the inner knot (J3) of A30, finding that the optical recombination line (ORL) 
emission came from a dense ($n{_{\rm He}}$ = 10${^4}$ cm$^{-3}$) and cold ($T_{\rm e}$ = 10${^3}$ K) core, 
whereas the collisionally excited line (CEL) 
emission would originate at a less dense ($n{_{\rm He}}$= 1700 cm$^{-3}$) but very hot ($T_{\rm e}$= 16\,000 K) 
envelope.  The 
long-standing ORL/CEL abundance discrepancy might find an explanation if density and chemical inhomogeneities 
such as those of J3 were ubiquitous in ionized regions. 

\subsection{Knots in Normal PNe}

An exciting discovery in NGC~5315 (a normal PN with an H-deficient central 
star) has been reported by Pottasch et al.\ (2002): the presence of two He-rich microstructures (of size 
$\sim$0.6 arcsec) located in a bipolar fashion in the inner 
nebula. Its origin is not known and further work, including measurement of their abundances and
kinematics, is required.
Perinotto et al.\ (2003) analyze the physico--chemical conditions of the FLIERS (and their antagonists, the so 
called SLOWERS) in NGC~7662. The previously reported N overabundance of the FLIERS (Balick et al.\ 1994) is not 
confirmed, although the analysis is limited by  uncertainties concerning the 
role of 
charge-exchange reactions in the derived abundances. More on this issue in the following section.

\subsection{Microstructures of NGC~7009}
Czysak \& Aller (1979) discovered a 
strong nitrogen overabundance in the outer knots (the ansae) of NGC~7009, a result  
confirmed by Balick et al.\ (1994), who found five times more N in the western ansa than at the rim; this was 
interpreted as the product of the recent high velocity ejection of N-rich material from the central star. 
Similar overabundances were also found in the low ionization knots of two other nebulae, NGC~6543 and NGC~6826, 
resulting in N/O values up to six times larger than in the cores. Ten years later, the origin of this 
N-richness and of the FLIERS themselves remains a mystery (Gon\c calves et al.\ 2001).
\begin{figure}
\epsscale{.90}
\plotone{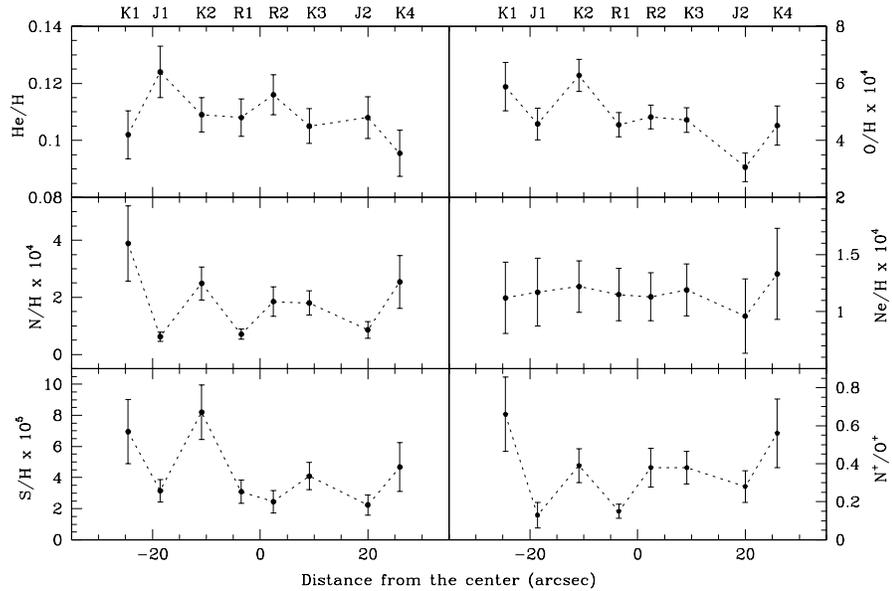}
\caption{Abundance profiles of NGC~7009 for He, O, N, Ne, S, and the N$^+$/O$^+$ 
ratio. 
The positions of the eight selected features are marked in the upper part of the plot. After Gon\c calves at 
al.\ 
(2003).} \label{T1.10-fig-1}
\end{figure} 

Gon\c calves et al.\ (2003) studied the rim, jets, and knots of NGC~7009 with deep optical spectra and a 
detailed 
analysis of the errors affecting both the physical parameters and the abundances. They find a moderate 
overabundance, 
of a factor of two, in N/H and N/O in the outer knots K1 and K4 (Figure~\ref{T1.10-fig-1}.
For an image of the nebula and its microstructures see Figure 2 in Gon\c calves's contribution in this volume). 
Obviously, this cast doubts on 
previous results and on the process itself for calculating abundances via the ICFs, but, if {\it {any}} 
overabundance  
prove to be real, then the problem of its origin would persist.

So is the nitrogen overabundance in FLIERS real or a conspiracy? Alexander \& Balick (1997) and 
Gruenwald \& Viegas (1998) showed that long-slit data may give spurious overabundances of N and other 
elements in the outer regions, and knots, of model PNe.  The main culprit is the different 
charge-exchange reaction rates of N and O, affecting the N$^+$/O$^+$ ratio mainly in the external 
low-ionization regions of the nebulae. As N$^+$/O$^+$ is usually assumed to be a good ICF for nitrogen, any 
extra quenching of O$^+$ with respect to N$^+$ caused by the charge-exchange reactions will translate into an N 
overabundance via the overestimated ICF.  
However,  it is not clear how close simple geometries and plane--parallel models like CLOUDY approach complex 
objects such as the microstructures of NGC~7009, nor why other elements such as Ne, Ar, S, etc., also 
expected to show spurious overabundances, do not do so.  This unresolved issue continues to prevent a robust 
interpretation of observed abundances in microstructures (Perinotto et al.\ 2003).

One approach would be to try to work with ICFs of order unity, i.e., observing the whole range of lines of the 
different ions, so avoiding the use of, in particular, the O$^+$ ionic abundance in the total N abundance.  
However, 
this implies getting adequate UV and IR data for the microstructures, which is not an easy task today. The 
alternative is to tailor-model individual cases with an appropriate 3D photoionization code that allows complex 
geometries and density distributions, and possible abundance variations. We have made a preliminary test of the 
role of charge-exchange reactions on measured abundances in NGC~7009 using MOCASSIN (Ercolano et al.\ 2003) 
with 
the following input parameters:  i) a central star of $T_{\rm eff}$ = 82\,000 K and luminosity 2500 $L_{\odot}$ 
at an 
assumed  distance of 1 kpc; ii) three constant density components:  R1, modeled as an  ellipsoidal shell with 
density 
4910 cm$^{-3}$; J1, as a cylindrical jet with 1155 cm$^{-3}$;  and K1, as a disk-shaped knot perpendicular to 
J1 with 2300 cm$^{-3}$; iii) temperatures are self-consistently derived in each region from modeling, and iv) 
abundances are those empirically derived for each component, except for K1 where N/H = 1.15 10$^{-4}$ is 
assumed (giving best fit for observed N lines in K1). The effect on the N/O abundance ratio of including 
({\it{on}} model) 
or excluding ({\it{off}} model) all charge-exchange reactions in the calculations is 
[N${^+}$/O${^+}$]$_{\rm off}$/[N${^+}$/O${^+}$]$_{\rm on}$ = 1.13, 1.16, and 1.18 for R1, J1, and K1, 
respectively.
It appears, therefore, that charge-exchange reactions are responsible for only a modest 
overestimation of N/O, of order 20\% in the external knots. Either the nitrogen enrichment at the knots is
real, or another factor is affecting the results. As discussed by Gon\c calves et al.\ (2003), neither 
collisional 
quenching of the [OII] lines nor strong shocks (enhancing the [NII] lines) seems at work at K1 and K4, and 
one is forced to accept that N/O can be really twice as large in these knots than at the rim in NGC~7009. The 
issue is currently being investigated with a more detailed model that includes the whole nebula.

\acknowledgements I thank R. L. M. Corradi and D. R. Gon\c calves for many lively discussions,
B. Balick, N. Soker and the APNIII organizing committee for their invitation, and M. Perinotto, A. Manchado,
M. A. Guerrero, E. Villaver, J. P. Phillips and F. Sabbadin for updating me on their recent work.  
Special thanks are due to S. R. Pottasch for sending me a lot of interesting data prior to publication.

%
%
%
%


\end{document}